\documentclass[letterpaper]{article} 
\usepackage{aaai2026}  
\usepackage{times}  
\usepackage{helvet}  
\usepackage{courier}  
\usepackage[hyphens]{url}  
\usepackage{graphicx} 
\usepackage{natbib}  
\usepackage{caption} 
\usepackage{amsmath}
\usepackage{amsfonts}
\usepackage{array}
\usepackage{booktabs}
\usepackage{enumitem}
\usepackage{subcaption}
\usepackage{tabularx}

\usepackage{algorithm}
\usepackage{algorithmic}

\usepackage{newfloat}
\usepackage{listings}
\DeclareCaptionStyle{ruled}{labelfont=normalfont,labelsep=colon,strut=off} 
\floatstyle{ruled}
\newfloat{listing}{tb}{lst}{}
\floatname{listing}{Listing}
\title{HiLoMix: Robust High- and Low-Frequency Graph Learning Framework \\ for Mixing Address Association}
\author{
    Xiaofan Tu, Tiantian Duan, Shuyi Miao, Hanwen Zhang, Yi Sun
}
\affiliations{

}

\usepackage{bibentry}

\begin{document}

\maketitle

\begin{abstract}
As mixing services are increasingly being exploited by malicious actors for illicit transactions, mixing address association has emerged as a critical research task. A range of approaches have been explored, with graph-based models standing out for their ability to capture structural patterns in transaction networks. However, these approaches face two main challenges: label noise and label scarcity, leading to suboptimal performance and limited generalization. To address these, we propose HiLoMix, a graph-based learning framework specifically designed for mixing address association. First, we construct the Heterogeneous Attributed Mixing Interaction Graph (HAMIG) to enrich the topological structure. Second, we introduce frequency-aware graph contrastive learning that captures complementary structural signals from high- and low-frequency graph views. Third, we employ weak supervised learning that assigns confidence-based weights to noisy labels. Then, we jointly train high-pass and low-pass GNNs using both unsupervised contrastive signals and confidence-based supervision to learn robust node representations. Finally, we adopt a stacking framework to fuse predictions from multiple heterogeneous models, further improving generalization and robustness. Experimental results demonstrate that HiLoMix outperforms existing methods in mixing address association. 
\end{abstract}


\section{Introduction}
\begin{figure}[tbp]
  \includegraphics[width=\linewidth]{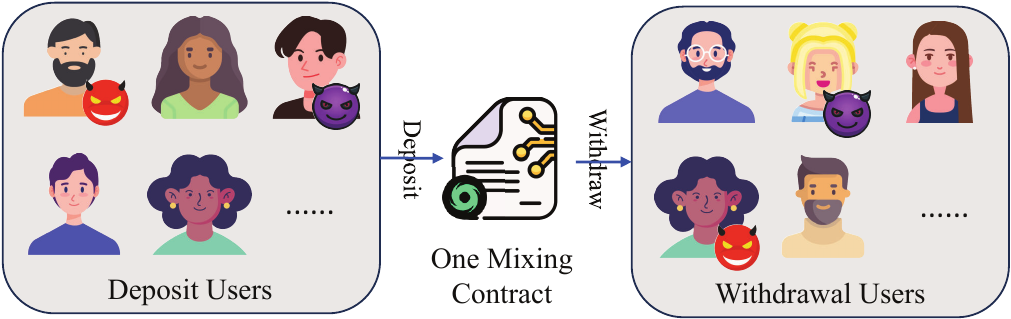}
  \caption{Mixing process of Tornado Cash. Characters marked with a devil icon represent illicit users. Two characters sharing the same icon indicate that they have transferred funds through mixer.}
  \label{fig:mixing}
\end{figure}

Web3.0 represents a paradigm shift toward a decentralized and user-centric internet, with Ethereum serving as its core infrastructure. Identities on Ethereum are represented by account addresses that are decoupled from real-world identities; however, this pseudonymity remains vulnerable to existing deanonymization attacks \cite{miao2025know, shen2021accurate, zhou2022behavior}. To mitigate these risks, users often turn to mixing services, among which \textbf{Tornado Cash} is the most widely used Ethereum mixer. As illustrated in Figure~\ref{fig:mixing}, it enables users to deposit assets into a funding pool and withdraw them without revealing which specific deposit it corresponds to, thus obscuring transaction trails and enhancing anonymity. However, Tornado Cash has also raised significant security concerns. According to Elliptic reports, Tornado Cash facilitated over \$ 1.5 billion in illicit transfers prior to being sanctioned by OFAC in August 2022. With the recent lifting of this sanction, its potential misuse for money laundering may escalate. These developments highlight the urgent need to deanonymize Tornado Cash by uncovering associations among addresses involved in mixing. 

Several approaches have been proposed to establish associations among mixing addresses, including heuristic and machine learning-based approaches. Heuristic approaches \cite{beres2021blockchain, wang2023zero} rely on empirical patterns and domain-specific assumptions, but struggle to detect malicious actors who deliberately evade detection and suffer from poor scalability. In contrast, machine learning approaches automate the association of mixing addresses by training predictive models that learn implicit transaction patterns from blockchain data, offering greater generalization and scalability. However, such approaches face two fundamental challenges in practice:
\begin{itemize}
    \item Label noise: As available labels are often derived from heuristic rules, they inevitably contain noise, introducing ambiguity during training and impairing the model’s generalization ability.
    \item Label scarcity: The limited availability of ground-truth associations prevents the model from establishing comprehensive decision boundaries, significantly degrading its performance and generalization ability.
\end{itemize}

\cite{hu2023bert4eth, hu2024zipzap} employed a pre-trained Transformer \cite{vaswani2017attention} to reduce reliance on labeled data. However, their approach overlooks the structural relationships among accounts, which are crucial for mixing address association, as true associations are often embedded in complex multi-hop interaction structures, which sequence modeling alone cannot capture. In contrast, \cite{beres2021blockchain, du2023breaking} modeled account interactions as graphs, enabling the capture of topological structures and interaction dependencies between accounts, and attempted to alleviate the label scarcity problem through techniques such as synthetic labeling. However, these challenges remain largely unresolved and existing methods still suffer from the following \textbf{limitations}:

\begin{enumerate}[label=\arabic*)]
  \item \textbf{Sparse graph structure.} The scarcity of associations results in highly sparse graph structures, leaving many nodes isolated and limiting the efficiency of message passing and feature aggregation.
  \item \textbf{Limited supervision.} Limited associations hinder models from learning comprehensive decision boundaries, leading to poor generalization performance in large-scale real-world transaction environments.
  \item \textbf{Overlooked label noise.} Heuristic labels are inherently noisy, which not only compromises training and evaluation, but also perturbs the graph structure, distorting connectivity patterns and neighborhood information, thereby impairing learning effectiveness.
\end{enumerate} 

To address the above limitations, we propose \textbf{HiLoMix}, a robust \underline{\textbf{Hi}}gh-frequency and \underline{\textbf{Lo}}w-frequency graph learning framework tailored for mixing address association. First, to alleviate graph sparsity, we construct the \textit{Heterogeneous Attributed Mixing Interaction Graph} (HAMIG), which encodes both account associations and transaction interactions as edges, enriching the graph topology and enhancing message propagation. Second, to mitigate limited supervision, we introduce a frequency-aware graph contrastive learning paradigm that captures complementary structural signals by contrasting node representations from high- and low-frequency views. Third, to handle label noise, we apply confidence-based label-weighted supervised learning, which assigns reliability-aware weights to noisy labels, enabling the model to leverage weak supervision without being dominated by noise. We jointly train a high-pass and low-pass GNN using both unsupervised contrastive signals and weighted supervision to learn robust node representations. Finally, we adopt the stacking framework that integrates the predictions of three heterogeneous base models, logistic regression, random forest, and multilayer perceptron, to further enhance generalization and robustness. 

Our main \textbf{contributions} are summarized as follows:
\begin{itemize}
    \item We propose HiLoMix, a robust high- and low-frequency graph learning framework tailored for the mixing address association task. 
    \item We adopt heterogeneous graph modeling, frequency-aware graph contrastive learning, and confidence-based label-weighted supervised learning to address the above limitations of existing approaches.
    \item HiLoMix achieves state-of-the-art performance, outperforming the best baseline by 5.69\%, 7.34\% and 15.61\% in $\text{F}_1$, AUC and MRR, respectively. Furthermore, we curate an up-to-date ground-truth dataset to facilitate future research on mixing address association.
\end{itemize}

\section{Background and Related Work}
\subsection{Ethereum and Tornado Cash}
Unlike Bitcoin, which supports only limited scripting capabilities for deterministic verification, Ethereum enables the deployment and execution of smart contracts through its Turing-complete Ethereum Virtual Machine (EVM). These smart contracts are immutable self-executing code deployed on-chain, capable of autonomously managing digital assets and enforcing application logic without centralized control. This decentralized computational infrastructure underpins a wide range of applications, including decentralized finance (DeFi), digital identity systems, decentralized autonomous organizations, and non-fungible tokens (NFTs), thereby fostering a trustless and transparent digital ecosystem.

Tornado Cash is a decentralized, non-custodial protocol designed to enhance transaction privacy on Ethereum by operating as a smart contract-based coin mixer. It employs Zero-Knowledge Succinct Non-Interactive Arguments of Knowledge (zk-SNARKs) to conceal the on-chain linkage between source and destination addresses, thereby significantly complicating efforts to trace the flow of funds. By cryptographically severing the deterministic connection between sender and receiver, Tornado Cash preserves transaction anonymity while maintaining on-chain verifiability.

\subsection{Summary of Existing Studies}
\textbf{Heuristic approaches} are rule-based approaches that infer mixing address associations from empirical patterns and domain-specific assumptions, rather than formal models or labeled data. \cite{hong2018practical} first explored such associations by linking addresses in the Helix Bitcoin mixer, based on the observation that user deposits were approximately equal to their withdrawals. \cite{beres2021blockchain, wang2023zero} extended these ideas by leveraging more transaction-level features, such as timing, distinctive gas prices, multi-denomination behavior, and TORN mining, to improve address linkage. Despite extensive exploration, heuristic approaches face inherent limitations. First, they rely on low-level transaction features (e.g. time, gas price) and require exhaustive pattern matching across all transactions, which limits scalability \cite{du2023breaking}. Second, rule-based approaches have limited detection capability, as only a small fraction of users match the predefined patterns. In many cases, they fail to identify malicious actors who actively evade detection and instead capture only inadvertent or careless users.

\textbf{Machine learning-based approaches} automate the association of mixing addresses by training predictive models that learn implicit transaction patterns from blockchain data. \cite{hu2023bert4eth} employed a pre-trained Transformer \cite{vaswani2017attention} to encode transaction sequences of Ethereum accounts, capturing temporal and contextual behavior patterns. Subsequently, \cite{hu2024zipzap} achieved both parameter and computational efficiency using frequency-aware compression and asymmetric training compared \cite{hu2023bert4eth}. However, their approaches overlook the structural relationships among accounts, which are crucial for mixing address association. \cite{du2023breaking} modeled the address interactions as a Mixing Interaction Graph (MIG) and formulated the task as a node-pair link prediction problem. By applying GraphSAGE \cite{hamilton2017inductive}, their approach captures the topological structures and interaction dependencies among accounts. Furthermore, \cite{beres2021blockchain} applied classic graph representation learning for address associations. \cite{zhong2024bitlink} developed a deep neural network to link disjoint Bitcoin address clusters in a self-supervised manner. Although these approaches are not specifically designed for mixing services, they can be readily adapted to our task due to the similarity in task formulation. However, machine learning models heavily rely on labeled data, which is unavailable for the mixing address association task. As a workaround, heuristic rules are employed to generate address associations but inevitably introduce noise, compromising both training and evaluation. To the best of our knowledge, the impact of such label noise has not been formally investigated in this context.

\section{Methodology}
\begin{figure*}[htbp]
  \centering
  \includegraphics[width=\linewidth]{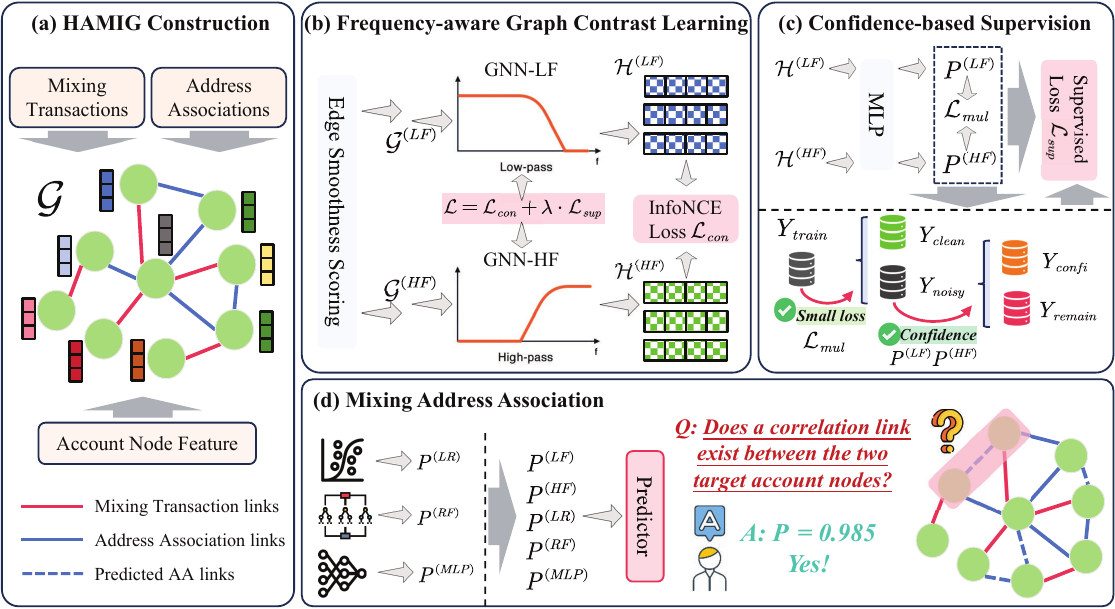}
  \caption{The overview framework of our HiLoMix.}
  \label{fig:pipeline}
\end{figure*}

\subsection{Overview}
As depicted in Figure~\ref{fig:pipeline}, HiLoMix comprises four key components: (1) \textbf{HAMIG construction}, which encodes diverse account interactions into a heterogeneous graph to enrich the topological structure; (2) \textbf{Frequency-aware graph contrast learning}, which contrasts node representations from high- and low-frequency views to capture robust structural patterns; (3) \textbf{Confidence-based supervision}, which assigns adaptive weights to noisy labels based on estimated reliability to mitigate supervision noise; (4) \textbf{Mixing address association}, which employs a stacking framework to integrate the knowledge of multiple models for final prediction. 

\subsection{HAMIG Construction}
\cite{du2023breaking} introduced the Mixing Interaction Graph (MIG) to capture the topological structure of account associations, where each account is modeled as a node and the address associations are represented as edges. Although effective in describing the basic interaction topology, MIG suffers from structural sparsity due to limited address associations, which restricts effective message propagation in graph-based learning. To address this limitation, we extend MIG by incorporating mixing transaction interactions as additional edges, thereby constructing a heterogeneous account interaction graph that significantly enriches the structural connectivity and semantic information of MIG. In addition, we extract a node feature matrix $\mathcal{X} \in \mathbb{R}^{n\times d}$, where $n$ denotes the number of account nodes and $d$ the feature dimension. Each feature vector encodes an account’s interaction behaviors with Tornado Cash mixing contracts, including its interaction frequency, timestamps, and gas price statistics. We define the resulting enhanced structure as the Heterogeneous Attributed Mixing Interaction Graph (HAMIG), which jointly captures both the topological dependencies among accounts and the attribute characteristics of individual nodes. Unlike MIG, HAMIG is formulated as an undirected heterogeneous graph $\mathcal{G} = (\mathcal{V}_a, \mathcal{V}_t, \mathcal{E}_{at}, \mathcal{E}_{aa}, \mathcal{X})$, where $\mathcal{V}_a$ denotes the set of account addresses interacting with Tornado Cash, $\mathcal{V}_t$ denotes the set of smart contracts within Tornado Cash, $\mathcal{E}_{at} = \{ (v_i, v_j) \mid v_i \in \mathcal{V}_a, v_j \in \mathcal{V}_t \}$ represents mixing transaction edges, $\mathcal{E}_{aa} = \{ (v_i, v_j) \mid v_i, v_j \in \mathcal{V}_a \}$ represents address association edges, and $\mathcal{X}$ is the node feature matrix.

\subsection{Frequency-aware Graph Contrast Learning} 
To mitigate insufficient supervision caused by label scarcity, we employ contrastive learning, which enables model training without ground-truth labels by leveraging instance-level similarity. Frequency-aware graph contrastive learning (GCL) offers a principled augmentation strategy by semantically meaningful and structurally coherent views based on the spectral characteristics of graph signals \cite{wan2024s3gcl}. However, directly applying biased-pass filters to the original graph signals may lead to entangled views with overlapping frequency components. To address this, inspired by \cite{wu2024robust}, we first decompose the graph signal into high- and low-frequency components before filtering. This separation enables each GNN branch to focus on a distinct frequency view, facilitating representation learning while minimizing cross-view interference. Specifically, we employ a multilayer perceptron (MLP) to estimate the edge smoothness probability $s_{ij}$ between node $v_i$ and $v_j$, where both node features are taken into account. Given the node feature matrix $\mathcal{H}_0 = \text{MLP}(\mathcal{X})$, the edge-wise smoothness is computed as follows: 
\begin{equation}
s_{ij} = \text{Sigmoid} \left( \text{Linear}\left(h_i \parallel h_j \right) + \text{Linear}\left(h_j \parallel h_i \right) \right)
\end{equation}
where $h_i$ indicates the feature vector of the $i$-th node, $\parallel$ denotes vector concatenation, and the sigmoid function ensures that $s_{ij}$ is bounded within $[0, 1]$. Based on estimated edge smoothness scores, HAMIG $\mathcal{G}$ is decomposed into two frequency-specific graph views: a low-frequency view $\mathcal{G}^{(LF)}$ and a high-frequency view $\mathcal{G}^{(HF)}$. Both views share the same node set, while the adjacency matrix $A$ is partitioned into $A^{(LF)}$ and $A^{(HF)}$ as follows:
\begin{equation}
\begin{aligned}
    A^{(LF)}_{ij} = s_{ij}&, \quad A^{(HF)}_{ij} = 1 - s_{ij} &\text{where } A_{ij} = 1 \\
    A^{(LF)}_{ij}& = A^{(HF)}_{ij} = 0 &\text{where } A_{ij} = 0
\end{aligned}
\end{equation}

To capture the complementary structural semantics embedded in different frequency bands of the graph signal, we design a dual-branch frequency-aware architecture to process low- and high-frequency components separately. This design is based on the principle that graph frequency characterizes the rate of variation in the node features across the graph topology: low-frequency components capture smooth, global structures, while high-frequency components highlight sharp local changes and structural anomalies.

For the low-frequency view $\mathcal{G}^{(LF)}$, we employ a low-pass graph neural network to propagate and smooth features across structurally similar neighborhoods, thereby capturing consistent global semantics.
\begin{equation}
    \mathcal{H}_l^{(LF)} = \left( I + \tilde{A}^{(LF)} \right) \mathcal{H}_{l-1}^{(LF)},\quad l \in \{1, ..., L\}
\end{equation}

In contrast, for the high-frequency view $\mathcal{G}^{(HF)}$, we utilize a high-pass GNN to preserve discriminative signals and amplify local heterogeneity based on the normalized graph Laplacian matrix.
\begin{equation}
    \mathcal{H}_l^{(HF)} = \left( I - \alpha \cdot \tilde{A}^{(HF)} \right) \mathcal{H}_{l-1}^{(HF)},\quad l \in \{1, ..., L\}
\end{equation}
Here, $\tilde{A}^{(LF)}$ and $\tilde{A}^{(HF)}$ denote the normalized adjacency matrices for the low- and high-frequency graph views, respectively; $\alpha$ controls the high-pass filter intensity; and $\mathcal{H}_L^{(LF)}$ and $\mathcal{H}_L^{(HF)}$ (i.e., $\mathcal{H}^{(LF)}$ and $\mathcal{H}^{(HF)}$) are the final node embeddings from the two views.

By applying low-pass and high-pass GNNs to the low- and high-frequency graph views, respectively, we obtain node representations that emphasize distinct frequency semantics. These representations, $\mathcal{H}^{(LF)}$ and $\mathcal{H}^{(HF)}$, are then contrasted using InfoNCE loss \cite{oord2018representation} to facilitate node representation learning.
{\fontsize{9}{0}
\begin{equation}
\mathcal{L}^{(LF)}\left(h_i^{(LF)}, h_i^{(HF)}\right) = -\log \frac{s\left(h_i^{(LF)}, h_i^{(HF)}\right)}{\sum_{j=1}^N s\left(h_i^{(LF)}, h_j^{(HF)}\right)}
\end{equation}
}
{\fontsize{9}{0}
\begin{equation}
\mathcal{L}^{(HF)}\left(h_i^{(HF)}, h_i^{(LF)}\right) = -\log \frac{s\left(h_i^{(HF)}, h_i^{(LF)}\right)}{\sum_{j=1}^N s\left(h_i^{(HF)}, h_j^{(LF)}\right)}
\end{equation}
}where {\fontsize{9}{0} $s\left(h_i^{(LF)}, h_i^{(HF)}\right)=\exp\left(\omega\left(h_i^{(LF)}, h_i^{(HF)}\right) / \tau\right)$}, with $\omega$ denoting cosine similarity and $\tau$ representing the temperature coefficient. This formulation encourages the model to produce semantically aligned embeddings across the two frequency-specific views. The overall contrastive objective is then defined as:
\begin{equation}
\mathcal{L}_{con} = \frac{1}{2|\mathcal{E}_{aa}|} \sum_{e \in \mathcal{E}_{aa}} \left[ \mathcal{L}^{(LF)} + \mathcal{L}^{(HF)} \right]
\end{equation}

\subsection{Confidence-based Supervision}
Although unsupervised paradigms such as contrastive learning help reduce dependence on labels and alleviate the impact of noise, they often fall short in performance compared to supervised methods. To address this limitation, we adopt a hybrid training strategy that integrates supervised learning signals to guide model optimization more effectively. Specifically, we compute the predicted probabilities of the association edge $e_{ij}$ from the two frequency-specific networks using the formulation $p_{ij} = \text{Linear}\left(h_i \parallel h_j\right)$. Based on these predictions, the mutual loss \cite{qian2023robust} for edge $e_{ij}$ is defined as:
\begin{align}
\mathcal{L}_{mul}^{ij} &= -y_{ij} \left[ \log \left(p_{ij,\; y_{ij}}^{(LF)}\right) + \log \left(p_{ij,\; y_{ij}}^{(HF)}\right) \right] \nonumber \\
&= -y_{ij} \log \left(p_{ij,\; y_{ij}}^{(LF)} \cdot p_{ij,\; y_{ij}}^{(HF)}\right)
\end{align}
Mutual loss gauges predictive confidence: $\mathcal{L}_{mul}^{ij}$ has a low value when both networks confidently and correctly predict the presence of edge $e_{ij}$. 

\subsubsection{\textbf{Clean Label Set Extraction}}
Deep neural networks exhibit a memorization effect, where clean and simple patterns are learned in the early stages of training, while noisy labels are gradually overfitted in later epochs \cite{arpit2017closer}. To adaptively manage label noise during training, we adopt the small-loss criterion \cite{han2018co}, which selects clean samples based on training losses. In epoch $t$, we define a threshold to distinguish clean from noisy labels as follows:
\begin{equation}
    \mathcal{L}_{thre}^{t} = \text{Percentile}\left(\mathcal{L}_{mul}^{ij}, 1 - \frac{t}{2\cdot T_{\text{max}}}\right)
\end{equation}
where $ T_{\text{max}} $ denotes the total number of training epochs, $ 0.5 \leq \left(1-\frac{t}{2\cdot T_{\text{max}}}\right) < 1 $ ensures that at least half of the labels are retained as clean, $\text{Percentile}\left(\mathcal{L}, p\right)$ returns the value below which $100*p$\% of the loss values in $\mathcal{L}$ fall. This dynamic threshold allows more samples to be included during the early stages for sufficient training and gradually reduces the number of selected instances as training progresses.

We define $\mathcal{L}_{thre}^{avg} = \text{Average} (\mathcal{L}_{mul}^{ij})$ to further ensure that small loss samples are clean regardless of their relative rank. Then clean and noisy label sets can be divided as follows:
\begin{equation}
\mathcal{E}_{cl} = \left\{ e_{ij} \in \mathcal{E}_{aa} \left| \mathcal{L}_{mul}^{ij} < \max \left( \mathcal{L}_{thre}^{t}, \mathcal{L}_{thre}^{avg} \right) \right. \right\}
\end{equation}

\subsubsection{\textbf{Confident Label Set Extraction}}
Given that $\mathcal{E}_{cl}$ denotes the clean label set, we define the remaining potentially noisy labels as $\mathcal{E}_{ns} = \mathcal{E}_{aa} \setminus \mathcal{E}_{cl}$. Inspired by \cite{qian2023robust}, we further extract a subset $\mathcal{E}_{cf} \subseteq \mathcal{E}_{ns}$, where two predictions, $p_{ij}^{(LF)}$ and $p_{ij}^{(HF)}$, confidently agree on the same prediction that differs from the observed label $y_{ij}$. Formally, this subset is characterized as follows:
\begin{equation}
c = \arg \max p_{ij}^{(LF)} = \arg \max p_{ij}^{(HF)} \neq y_{ij}
\end{equation}
\begin{equation}
\mathcal{E}_{cf} = \left\{ e_{ij} \in \mathcal{E}_{ns} \, \middle| \, \mu(e_{ij}) > Thre =1 - \frac{t}{2 \cdot T_{\text{max}}} \right\}
\end{equation}
where the confidence score $\mu(e_{ij})$ is defined as $\mu(e_{ij}) = \sqrt{p_{ij, c}^{(LF)} \cdot p_{ij, c}^{(HF)}}$. A higher value of $\mu(e_{ij})$ suggests a greater likelihood that $e_{ij}$ is mislabeled, yet correctly predicted. The threshold $Thre$ is designed to be dynamic: stricter in the early stages of training and gradually relaxed as the training progresses. 

Given the previously defined sets $\mathcal{E}_{cl}$ and $\mathcal{E}_{cf}$, we denote the remaining labels as $\mathcal{E}_{re}$, from which no useful information can be reliably extracted.
\begin{equation}
\mathcal{E}_{re} = \mathcal{E}_{aa} \setminus \mathcal{E}_{cl} \setminus \mathcal{E}_{cf}
\end{equation}

After partitioning the labels into $\mathcal{E}{cl}$, $\mathcal{E}{cf}$, and $\mathcal{E}_{re}$, the confidence-based supervision loss for mixing address association is defined as follows:
{\fontsize{9}{0}
\begin{equation}
\mathcal{L}_{\mathrm{sup}}
= -\frac{1}{|\mathcal{E}_{aa}|}\sum_{e_{ij}\in\mathcal{E}_{aa}}
\xi(e_{ij})\big( \log p^{(LF)}_{ij,\;\hat y_{ij}} + \log p^{(HF)}_{ij,\;\hat y_{ij}} \big)
\end{equation}}
where
\begin{equation}
\left\{
\begin{aligned}
\xi(e_{ij})   &= 1,      &\quad \hat{y}_{ij} &= y_{ij}, &\quad &\text{if } e_{ij} \in \mathcal{E}_{cl} \\
\xi(e_{ij})   &= \mu(i), &\quad \hat{y}_{ij} &= c,      &\quad &\text{if } e_{ij} \in \mathcal{E}_{cf} \\
\xi(e_{ij})   &= 0.5,    &\quad \hat{y}_{ij} &= y_{ij}, &\quad &\text{if } e_{ij} \in \mathcal{E}_{re}
\end{aligned}
\right.
\end{equation}
where $\xi(\cdot)$ serves as a confidence score, assigning each label a different weight to guide the model towards more reliable supervision. Finally, the total loss used to update both the high-pass and low-pass GNNs is formulated as follows:
\begin{equation}
    \mathcal{L} = \mathcal{L}_{con} + \lambda \cdot \mathcal{L}_{sup}
\end{equation}

\subsection{Mixing Address Association}
After training the high-pass and low-pass graph neural networks, instead of simply selecting the better one or applying confidence-based voting, we adopt the stacking method \cite{wolpert1992stacked} to integrate their knowledge for the final prediction. Stacking aims to exploit the strengths of multiple base models by training a meta-learner on their predictions. In general, heterogeneous models, differing in learning algorithms, hypothesis spaces, or inductive biases, promote diversity and enhance ensemble performance. However, in our case, high-pass and low-pass GNNs are inherently homogeneous, limiting their ability to provide complementary information and even resulting in performance degeneration, as shown in Table~\ref{tab:ablation}. To overcome this limitation, we introduce three additional heterogeneous models, random forest, logistic regression, and multilayer perceptron, into the stacking ensemble, enriching it with diverse decision boundaries and representational capabilities. 

\section{Experiments}
\subsection{Experiment Setup}
\subsubsection{\textit{Dataset:}} 
We collect 554,589 raw Tornado Cash mixing transactions spanning from 2019-12-16 to 2025-01-17. After data cleaning and preprocessing, 371,181 valid transactions are retained. Following the heuristic rules proposed in \cite{wu2022tutela, wang2023zero}, we identify 4,074 unique address associations as labels. By modeling these associations and transaction interactions as edges, we construct HAMIG comprising 106,982 account nodes and 375,255 edges of two types. 

Although our dataset is constructed solely from Tornado Cash, we note that this does not compromise generality. Tornado Cash remains the largest and most representative Ethereum-based mixer, accounting for over 95\% of the total transaction volume and total value locked (TVL) across all Ethereum mixing services according to \cite{akopovadetecting}. Therefore, focusing on Tornado Cash already captures the predominant behavioral and structural patterns of Ethereum-based mixing activities. 

\begin{table*}[htbp]
\centering
  \begin{tabular}{@{}lcccccc@{}}
    \toprule
    Methods & $\text{F}_1$ & AUC & MRR & Hits@3 & Hits@5 & Hits@10 \\
    \midrule
    DeepWalk & 0.6844$\pm$0.0030 & 0.7078$\pm$0.0067 & 0.0866$\pm$0.0056 & 0.0568$\pm$0.0081 & 0.0932$\pm$0.0043 & 0.1940$\pm$0.0057 \\
    Node2Vec & 0.6970$\pm$0.0040 & 0.7251$\pm$0.0033 & 0.0850$\pm$0.0027 & 0.0553$\pm$0.0046 & 0.0926$\pm$0.0079 & 0.1870$\pm$0.0019 \\
    \midrule
    GCN & 0.7639$\pm$0.0035 & 0.7795$\pm$0.0063 & 0.1011$\pm$0.0090 & 0.0542$\pm$0.0114 & 0.1233$\pm$0.0239 & 0.3151$\pm$0.0369 \\
    GAT & 0.7674$\pm$0.0064 & 0.7828$\pm$0.0090 & 0.1025$\pm$0.0047 & 0.0569$\pm$0.0054 & 0.1217$\pm$0.0106 & 0.3179$\pm$0.0093 \\
    GIN & 0.7869$\pm$0.0036 & 0.8167$\pm$0.0068 & 0.1415$\pm$0.0096 & 0.1143$\pm$0.0135 & 0.1900$\pm$0.0210 & 0.3802$\pm$0.0327  \\
    GraphSAGE & 0.7867$\pm$0.0028 & 0.8138$\pm$0.0022 & 0.1187$\pm$0.0127 & 0.0794$\pm$0.0178 & 0.1535$\pm$0.0235 & 0.3568$\pm$0.0211 \\
    \midrule
    Co-teaching & 0.7859$\pm$0.0057 & 0.8011$\pm$0.0028 & 0.0961$\pm$0.0106 & 0.0460$\pm$0.0160 & 0.0927$\pm$0.0255 & 0.2578$\pm$0.0557 \\
    Co-teaching+ & 0.7690$\pm$0.0014 & 0.8230$\pm$0.0040 & 0.2712$\pm$0.0025 & 0.2994$\pm$0.0007 & 0.4281$\pm$0.0036 & 0.6169$\pm$0.0026 \\
    NRGL & 0.7866$\pm$0.0058 & \underline{0.8513$\pm$0.0029} & \underline{0.3989$\pm$0.0058} & \underline{0.4784$\pm$0.0079} & \underline{0.5896$\pm$0.0078} & \underline{0.7263$\pm$0.0038} \\
    \midrule
    BERT4ETH & 0.6990$\pm$0.0083 & 0.7897$\pm$0.0215 & 0.3366$\pm$0.0656 & 0.3878$\pm$0.0872 & 0.4694$\pm$0.0991 & 0.5714$\pm$0.0949 \\
    MixBroker & \underline{0.7930$\pm$0.0024} & 0.8182$\pm$0.0049 & 0.1062$\pm$0.0089 & 0.0585$\pm$0.0117 & 0.1231$\pm$0.0170 & 0.3323$\pm$0.0292 \\
    HiLoMix & \textbf{0.8382$\pm$0.0044} & \textbf{0.9137$\pm$0.0033} & \textbf{0.4612$\pm$0.0087} & \textbf{0.5173$\pm$0.0086} & \textbf{0.6556$\pm$0.0114} & \textbf{0.8037$\pm$0.0111} \\
    \midrule
    \%Improv. & 5.69\% & 7.34\% & 15.61\% & 8.13\% & 11.18\% & 10.65\% \\
  \bottomrule
\end{tabular}
\caption{Overall Performance Comparison. \textbf{Bold} and \underline{underline} represent the best and second best performance, respectively. We repeat the experiment with 3 random seeds and report the average metrics with standard deviation.} 
\label{tab:comparison}
\end{table*}

\subsubsection{\textit{Evaluation metrics:}} 
To comprehensively evaluate the performance of HiLoMix, we adopt four metrics across two categories: classification and ranking. Specifically, F1-score and AUC are used to evaluate classification performance, while MRR and Hits@K assess ranking effectiveness. The formal definitions of MRR and Hits@K are provided:
\begin{equation}
\mathrm{MRR} = \frac{1}{N} \sum_{i=1}^{N} \frac{1}{\mathrm{rank}_i}
\end{equation}
\begin{equation}
\mathrm{Hits@K} = \frac{1}{N} \sum_{i=1}^{N} \mathbb{I}(\mathrm{rank}_i \le K)
\end{equation}
where $N$ is the number of positive samples, $\mathrm{rank}_i$ denotes the ranking position of the $i$-th positive sample in the candidate list based on predicted probabilities, and $\mathbb{I}(\cdot)$ is the indicator function.

\subsubsection{\textit{Baselines:}} 
We compare HiLoMix with 11 baseline models, grouped into three categories: (1) Graph representation learning methods: DeepWalk \cite{perozzi2014deepwalk}, Node2Vec \cite{grover2016node2vec}, GCN \cite{kipf2016semi}, GAT \cite{velickovic2017graph}, GIN \cite{xu2018powerful}, and GraphSAGE \cite{hamilton2017inductive}. Since the mixing address association is formulated as a node-pair link prediction task in HAMIG, classical graph representation learning models serve as strong baselines. (2) Robust learning methods: Co-teaching \cite{han2018co}, Co-teaching+ \cite{yu2019does} and NRGL \cite{wu2024robust}. Given the central limitation of label noise in our task, we include both classical and state-of-the-art robust learning approaches for comparison. (3) Mixing address association methods: BERT4ETH \cite{hu2023bert4eth} and MixBroker \cite{du2023breaking}. These methods are designed specifically for the mixing address association task.

\subsubsection{\textit{Implementation Details:}}
All implementations are conducted using PyTorch 2.3.1 and DGL 2.4.0. The model is trained for 50 epochs with a batch size of 128, using the Adam optimizer with a learning rate of 0.003. All experiments are performed on a single NVIDIA A100 GPU. For contrastive loss, we use a batch of 128 nodes and set the temperature coefficient at 0.5. For MRR and Hits@K metrics, 50 negative samples are generated for each positive instance. All baselines are implemented following their original papers and fine-tuned to achieve optimal performance.

\subsection{Performance Comparison}
We conduct a systematic comparison between HiLoMix and the baseline methods. The results of the mixing address association performance are presented in Table~\ref{tab:comparison}. The improvement (\%Improv.) is calculated as the relative gain over the second-best model.

\textbf{Mixing address association under the classification setting.} HiLoMix achieves an AUC of 0.9137 and a $\text{F}_1$ score of 0.8382, outperforming all baselines with relative improvements ranging from 7.34\% to 29.09\% in AUC and 5.69\% to 22.47\% in $\text{F}_1$, demonstrating its effectiveness in the classification setting. Classic graph representation learning methods perform the worst, primarily due to their vulnerability to label noise and graph sparsity resulting from scarce associations. NRGL achieves the second highest AUC, validating the effectiveness of explicit label noise handling. MixBroker attains the second-best $\text{F}_1$ score, benefiting from task-specific graph modeling that captures user interaction structures. In contrast, BERT4ETH underperforms standard GNNs, highlighting the importance of incorporating structural information in address association.

\textbf{Mixing address association under the ranking setting.} HiLoMix outperforms the second-best model by 15.61\%, 8.13\%, 11.18\% and 10.65\% in MRR, Hits@3, Hits@5 and Hits@10, respectively. These results indicate that HiLoMix effectively ranks true address associations near the top among all candidates, demonstrating a strong generalization to unseen links. In line with the classification results, the classic graph representation learning methods perform the worst. NRGL ranks second across all ranking metrics, further validating the effectiveness of robust learning strategies for mixing address association task. BERT4ETH leverages sequence modeling to generate expressive account representations; however, its disregard for structural dependencies leads to suboptimal performance. MixBroker, while incorporating user interaction structures through graph modeling, suffers from graph sparsity and structural perturbations.

\subsection{Ablation Study}
To systematically validate the effectiveness of each component in HiLoMix, we conduct ablation studies by removing or modifying specific modules. Specifically, ``w/o hetero models" excludes all additional heterogeneous base models from the stacking ensemble. ``w/o HiLo GNNs" removes both high-pass and low-pass GNNs from the ensemble. ``w/o Hi GNN" and ``w/o Lo GNN" remove high-pass and low-pass GNN, respectively. ``w/o hetero graph" removes all mixing transaction edges, retaining only address association edges in HAMIG. ``w/o HiLo learning" replaces the high- and low-pass GNNs with two GCNs. ``w/o label division" disables the confidence-based label division mechanism. ``w/o stacking" discards the stacking framework and instead selects the better result between the high- and low-pass GNNs. 

\begin{table}[tbp]
  \centering
  \begin{tabular}{lccc}
    \toprule
    Methods & $\text{F}_1$ & AUC & MRR\\
    \midrule
    w/o hetero models & 0.7393 & 0.7948 & 0.2977 \\
    w/o HiLo GNNs & 0.8075 & 0.8761 & 0.3507 \\
    w/o Hi GNN & 0.8169 & 0.8977 & 0.4202 \\
    w/o Lo GNN & 0.8286 & 0.9046 & 0.4129 \\
    w/o hetero graph & 0.8158 & 0.8931 & 0.4419 \\
    w/o HiLo learning & 0.8083 & 0.8911 & 0.4228 \\
    w/o label division & 0.8238 & 0.8938 & 0.4260 \\
    w/o stacking & 0.7882 & 0.8547 & 0.4061 \\
    \midrule
    HiLoMix & \textbf{0.8382} & \textbf{0.9137} & \textbf{0.4612} \\
  \bottomrule
\end{tabular}
\caption{Ablation Study Results.}
\label{tab:ablation}
\end{table}

As shown in Table~\ref{tab:ablation}, removing any component results in a consistent drop in both classification and ranking metrics, confirming the complementary contributions of all modules. Among all variants, excluding heterogeneous base models leads to the greatest degradation, with $\text{F}_1$ decreasing by 11.80\% and MRR by 35.45\%. Although the drop is substantial, this result is expected: in the \textit{w/o hetero models} setting, the high-pass and low-pass GNNs are inherently homogeneous, limiting their ability to provide complementary information. In contrast, noticeable drops occur when the core GNNs are removed from the stacking ensemble, confirming their pivotal role in HiLoMix’s effectiveness. The remaining variants yield moderate performance decreases, further validating the necessity of each component. These smaller drops also demonstrate the robustness of HiLoMix and indicate that the stacking framework effectively integrates noise-resistant and complementary knowledge from heterogeneous base models, maintaining stable performance even when certain components are weakened.

\subsection{Feature Importance Evaluation}
We conduct a comparative study across seven feature variants by progressively removing less important dimensions. Additionally, we incorporate node features from MixBroker for further comparison. As shown in Table~\ref{tab:feature_cmp}, HiLoMix achieves the best performance across all metrics when using our 197-dimensional account node features, outperforming the second-best features, MixBroker features. Although the 221-dimensional feature set contains more attributes, its performance is inferior to that of the 197-dimensional set due to reduced interpretability. These results suggest that our 197-dimensional features not only enhance model explainability but also reduce HAMIG's memory footprint without compromising performance.

\begin{table}[tbp]
  \centering
  \begin{tabular}{lccc}
    \toprule
    Methods & $\text{F}_1$ & AUC & MRR \\
    \midrule
    28-Dim Feature & 0.8092 & 0.8780 & 0.3834 \\
    53-Dim Feature & 0.7984 & 0.8686 & 0.3691 \\
    105-Dim Feature & 0.8131 & 0.8761 & 0.3926 \\
    141-Dim Feature & 0.8067 & 0.8861 & 0.4261 \\
    167-Dim Feature & 0.8208 & 0.8957 & 0.4249 \\
    197-Dim Feature & \textbf{0.8382} & \textbf{0.9137} & \textbf{0.4612} \\
    221-Dim Feature & 0.8183 & 0.8948 & 0.4408 \\
    MixBroker Feature & 0.8248 & 0.9025 & 0.4423 \\
  \bottomrule
\end{tabular}
\caption{Performance Comparison of Different Features.}
\label{tab:feature_cmp}
\end{table}

\subsection{Hyper-parameter Sensitivity Analysis}
\begin{figure}[htbp]
  \centering
  \includegraphics[width=0.9\linewidth]{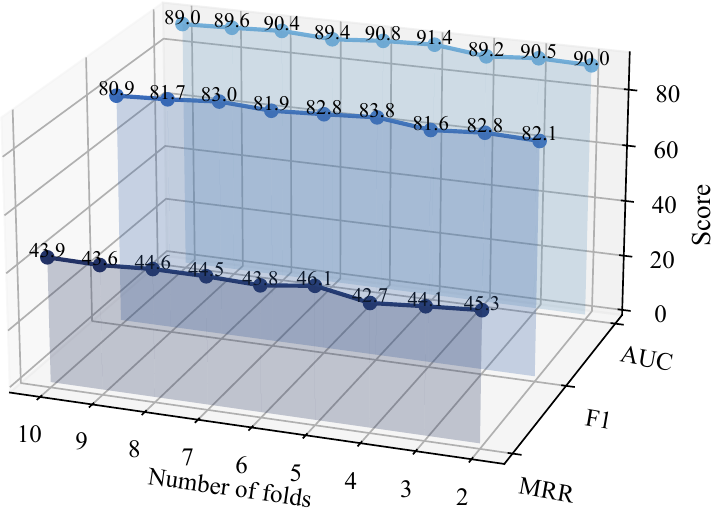}
  \caption{Performance of HiLoMix when training the base models in the stacking framework using different K of K-fold cross-validation.}
  \label{fig:fold}
\end{figure}

We perform a sensitivity analysis on key hyper-parameters in HiLoMix. We begin by evaluating model stability with respect to the number of folds used in cross-validation within the stacking ensemble. As shown in Figure~\ref{fig:fold}, HiLoMix performance remains generally stable across different fold settings. The best overall performance is observed with 5-fold cross-validation, which is therefore adopted as the default setting. Both fewer and greater numbers of folds lead to slight performance degradation, potentially due to overfitting or unstable ensemble integration. 
\begin{figure}[htbp]
    \centering
    \begin{subfigure}[b]{0.45\linewidth}
        \includegraphics[width=\linewidth]{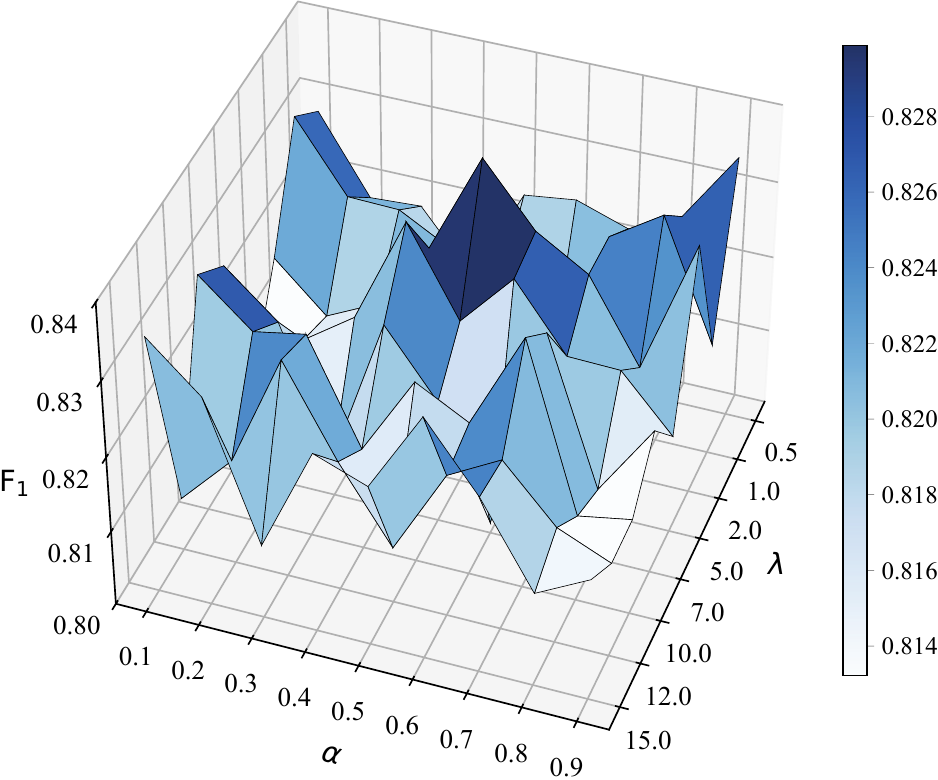}
        \caption{$\text{F}_1$ metric.}
    \end{subfigure}
    \hspace{0.05\linewidth}
    \begin{subfigure}[b]{0.45\linewidth}
        \includegraphics[width=\linewidth]{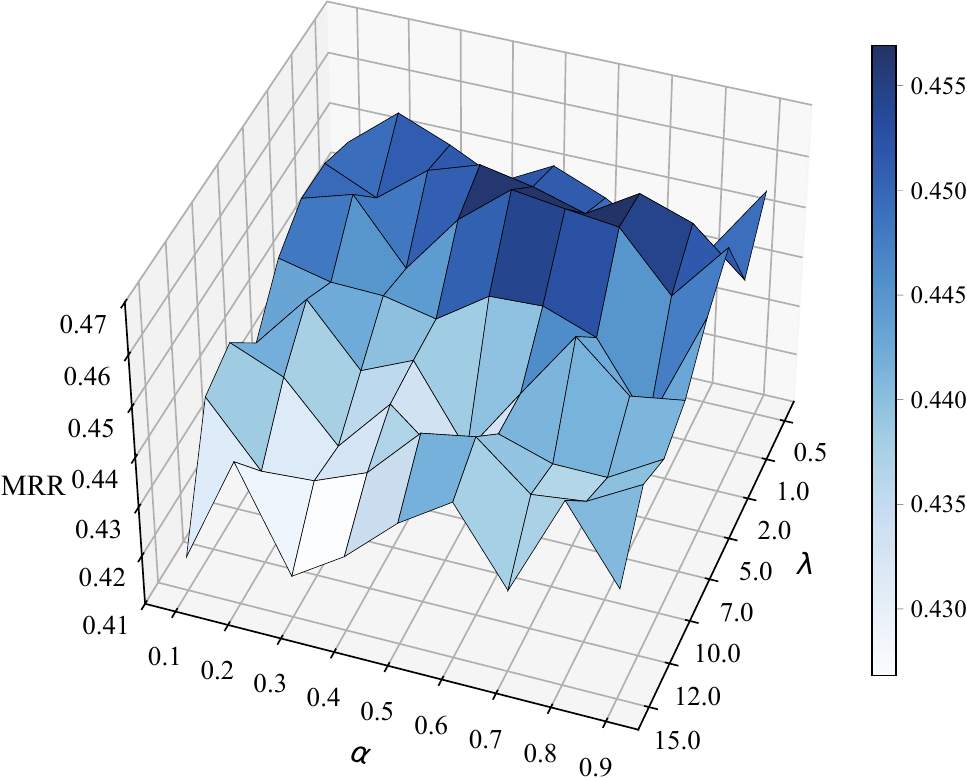}
        \caption{MRR metric.}
    \end{subfigure}
    \caption{Performance of HiLoMix under different settings of hyperparameters $\alpha$ and $\lambda$.}
    \label{fig:alpha-lambda}
\end{figure}

We further investigate two additional hyper-parameters: loss weighting $\lambda$, which balances the supervised and contrastive learning objectives, and the high-pass filter strength $\alpha$, which modulates the emphasis on high-frequency components. As illustrated in Figure~\ref{fig:alpha-lambda}, HiLoMix demonstrates robust performance across a wide range of $\lambda$ and $\alpha$ values. Both metrics peak around $\lambda = 2$ and $\alpha = 0.5$, suggesting that a moderate balance between the supervised and contrastive learning objectives and the well-calibrated filter strength are critical for optimal performance. 

\subsection{Efficiency Analysis}
Although HiLoMix introduces additional components compared to simple baselines, most of them are computationally lightweight and do not incur substantial overhead. We compare HiLoMix with representative baselines, i.e., the best-performing model from each baseline category. The average training time per epoch is reported in Table~\ref{tab:efficiency}.
\begin{table}[htbp]
\centering
\begin{tabular}{lcc}
\toprule
Method & Training time (s) \\
\midrule
HiLoMix     & 6.91  \\
GIN         & 0.26  \\
NRGL        & 15.27 \\
MixBroker   & 0.26  \\
\bottomrule
\end{tabular}
\caption{Comparison of training efficiency.}
\label{tab:efficiency}
\end{table}

As shown in Table~\ref{tab:efficiency}, HiLoMix incurs moderate additional computation compared to simple GNNs but remains substantially faster than NRGL. In contrast, DeepWalk/Node2Vec and BERT4ETH require several hours for random walk generation or pretraining to obtain node embeddings, making them far less efficient than our model. All models in our experiments, including HiLoMix and the baselines, complete inference on the entire test set in under 0.01 seconds. In general, HiLoMix achieves a favorable balance between performance and computational cost.

\subsection{Structural and Label Dynamics Analysis}
We quantitatively compare MIG and HAMIG in terms of key structural properties. As shown in Table~\ref{tab:stats}, HAMIG effectively alleviates the structural sparsity of MIG, increasing the average node degree by approximately 45$\times$, and also improving the clustering coefficient and graph density by several orders of magnitude. This enriched topology facilitates more efficient message propagation among nodes.
\begin{table}[tbp]
\centering
\begin{tabular}{lccc}
\toprule
Metric & MIG & HAMIG \\
\midrule
Average node degree & 0.0565 & 2.561 \\
Average clustering coefficient & 0.0004 & 0.0390 \\
Graph density & 4.93e-7 & 2.23e-5 \\
\bottomrule
\end{tabular}
\caption{Comparison of graph structural properties between MIG and HAMIG.}
\label{tab:stats}
\end{table}

\begin{figure}[tbp]
  \centering
  \includegraphics[width=\linewidth]{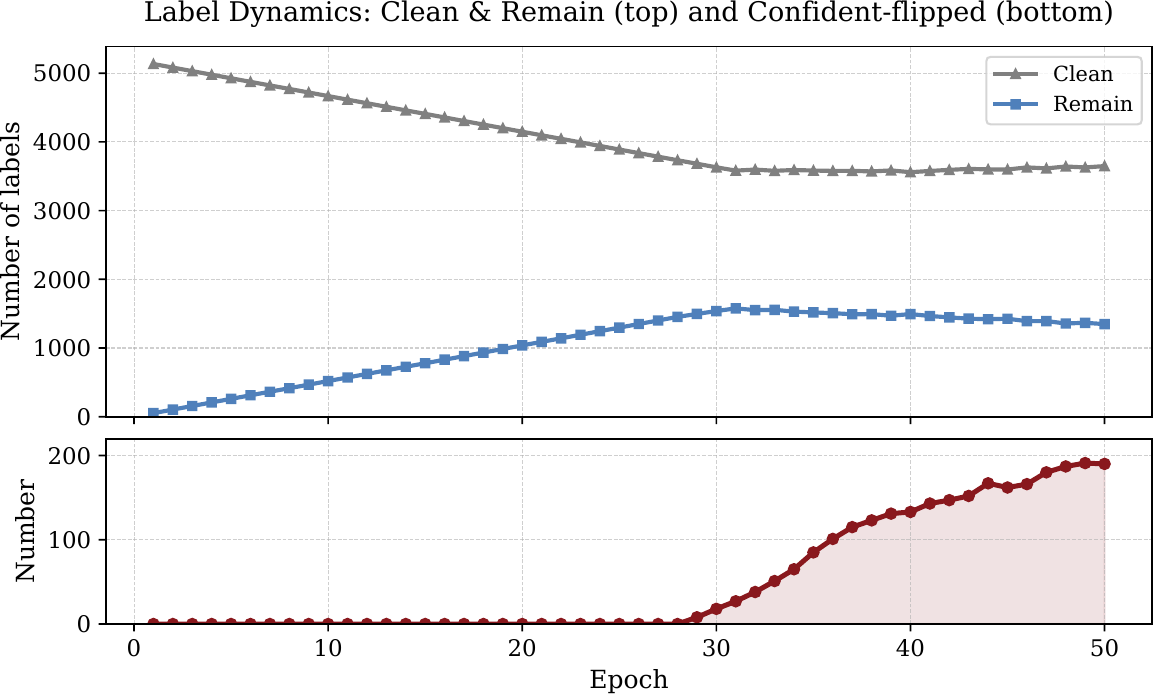}
  \caption{Dynamics of three label sets during training.}
  \label{fig:label}
\end{figure}

We track the number of \textit{clean}, \textit{confident-flipped}, and \textit{remaining} labels across epochs, as illustrated in Figure~\ref{fig:label}.
The evolution process can be roughly divided into two stages: early training (0–30 epochs) and later training (30–50 epochs). During the early stage, the number of clean labels gradually decreases, while remaining labels increase, as the model transitions from fitting clean samples to handling potentially noisy ones. At this point, confident-flipped labels remain nearly zero, indicating that the model has not yet learned sufficiently reliable structural patterns. In the later stage, the number of confident-flipped labels steadily increases, showing that the model begins to capture meaningful structural dependencies and distinguish true associations from incorrect ones. Meanwhile, clean labels stabilize, and the model progressively filters mislabeled samples from the remaining set, demonstrating improved robustness to noisy supervision. This dynamic evolution process confirms that confidence-based supervision effectively governs label noise and guides the model toward reliable learning.

\section{Conclusion}
In this work, we propose HiLoMix, a novel graph-based learning framework tailored for mixing address association. By modeling diverse account interactions in Tornado Cash as HAMIG, we effectively address the issue of graph sparsity. To alleviate the limited supervision and label noise problem, we introduce frequency-aware graph contrastive learning alongside confidence-based label-weighted supervised learning, jointly optimizing high-pass and low-pass GNNs for robust node representations. Extensive experiments demonstrate that HiLoMix consistently outperforms existing approaches on the mixing address association task. Furthermore, we curate an up-to-date ground-truth dataset to support and facilitate future research in this field.

\section*{Acknowledgements}
This work was supported by Beijing Advanced Innovation Center for Future Blockchain and Privacy Computing and the National Natural Science Foundation of China (U22B2032).

\bibliography{aaai2026}

@INPROCEEDINGS{miao2025know,
author={Miao, Shuyi and Qiu, Wangjie and Zheng, Hongwei and Zhang, Qinnan and Tu, Xiaofan and Liu, Xunan and Liu, Yang and Dong, Jin and Zheng, Zhiming},
booktitle = { 2025 IEEE 41st International Conference on Data Engineering (ICDE) },
title = { Know Your Account: Double Graph Inference-Based Account De-Anonymization on Ethereum },
year = {2025},
pages = {1305-1319},
}

@ARTICLE{shen2021accurate,
  author={Shen, Meng and Zhang, Jinpeng and Zhu, Liehuang and Xu, Ke and Du, Xiaojiang},
  journal={IEEE Transactions on Information Forensics and Security}, 
  title={Accurate Decentralized Application Identification via Encrypted Traffic Analysis Using Graph Neural Networks}, 
  year={2021},
  volume={16},
  pages={2367-2380},
}

@ARTICLE{zhou2022behavior,
  author={Zhou, Jiajun and Hu, Chenkai and Chi, Jianlei and Wu, Jiajing and Shen, Meng and Xuan, Qi},
  journal={IEEE Transactions on Information Forensics and Security}, 
  title={Behavior-Aware Account De-Anonymization on Ethereum Interaction Graph}, 
  year={2022},
  volume={17},
  pages={3433-3448},
}

@inproceedings{wang2023zero,
author = {Wang, Zhipeng and Chaliasos, Stefanos and Qin, Kaihua and Zhou, Liyi and Gao, Lifeng and Berrang, Pascal and Livshits, Benjamin and Gervais, Arthur},
title = {On How Zero-Knowledge Proof Blockchain Mixers Improve, and Worsen User Privacy},
year = {2023},
booktitle = {Proceedings of the ACM Web Conference 2023},
pages = {2022–2032},
}

@article{wu2022tutela,
  title={Tutela: An open-source tool for assessing user-privacy on ethereum and tornado cash},
  author={Wu, Mike and McTighe, Will and Wang, Kaili and Seres, Istvan A and Bax, Nick and Puebla, Manuel and Mendez, Mariano and Carrone, Federico and De Mattey, Tom{\'a}s and Demaestri, Herman O and others},
  journal={arXiv preprint arXiv:2201.06811},
  year={2022}
}

@article{du2023breaking,
author = {Du, Hanbiao and Che, Zheng and Shen, Meng and Zhu, Liehuang and Hu, Jiankun},
title = {Breaking the Anonymity of Ethereum Mixing Services Using Graph Feature Learning},
year = {2024},
issue_date = {2024},
volume = {19},
journal = {Trans. Info. For. Sec.},
pages = {616–631},
}

@inproceedings{hu2023bert4eth,
author = {Hu, Sihao and Zhang, Zhen and Luo, Bingqiao and Lu, Shengliang and He, Bingsheng and Liu, Ling},
title = {BERT4ETH: A Pre-trained Transformer for Ethereum Fraud Detection},
year = {2023},
booktitle = {Proceedings of the ACM Web Conference 2023},
pages = {2189–2197},
}

@inproceedings{vaswani2017attention,
author = {Vaswani, Ashish and Shazeer, Noam and Parmar, Niki and Uszkoreit, Jakob and Jones, Llion and Gomez, Aidan N. and Kaiser, \L{}ukasz and Polosukhin, Illia},
title = {Attention is all you need},
year = {2017},
booktitle = {Proceedings of the 31st International Conference on Neural Information Processing Systems},
pages = {6000–6010},
}

@INPROCEEDINGS{beres2021blockchain,
  author={Béres, Ferenc and Seres, István A. and Benczúr, András A. and Quintyne-Collins, Mikerah},
  booktitle={2021 IEEE International Conference on Decentralized Applications and Infrastructures (DAPPS)}, 
  title={Blockchain is Watching You: Profiling and Deanonymizing Ethereum Users}, 
  year={2021},
  pages={69-78},
}

@inproceedings{hamilton2017inductive,
author = {Hamilton, William L. and Ying, Rex and Leskovec, Jure},
title = {Inductive representation learning on large graphs},
year = {2017},
booktitle = {Proceedings of the 31st International Conference on Neural Information Processing Systems},
pages = {1025–1035},
}

@inproceedings{hu2024zipzap,
author = {Hu, Sihao and Huang, Tiansheng and Chow, Ka-Ho and Wei, Wenqi and Wu, Yanzhao and Liu, Ling},
title = {ZipZap: Efficient Training of Language Models for Large-Scale Fraud Detection on Blockchain},
year = {2024},
booktitle = {Proceedings of the ACM Web Conference 2024},
pages = {2807–2816},
}

@inproceedings{hong2018practical,
author = {Hong, Younggee and Kwon, Hyunsoo and Lee, Jihwan and Hur, Junbeom},
title = {A Practical De-mixing Algorithm for Bitcoin Mixing Services},
year = {2018},
booktitle = {Proceedings of the 2nd ACM Workshop on Blockchains, Cryptocurrencies, and Contracts},
pages = {15–20},
}

@inproceedings{zhong2024bitlink,
author = {Zhong, Sheng and Mueen, Abdullah},
title = {BitLINK: Temporal Linkage of Address Clusters in Bitcoin Blockchain},
year = {2024},
booktitle = {Proceedings of the 30th ACM SIGKDD Conference on Knowledge Discovery and Data Mining},
pages = {4583–4594},
}

@misc{oord2018representation,
      title={Representation Learning with Contrastive Predictive Coding}, 
      author={Aaron van den Oord and Yazhe Li and Oriol Vinyals},
      year={2019},
      eprint={1807.03748},
      archivePrefix={arXiv},
}

@inproceedings{wan2024s3gcl,
author = {Wan, Guancheng and Tian, Yijun and Huang, Wenke and Chawla, Nitesh V and Ye, Mang},
title = {S3GCL: spectral, swift, spatial graph contrastive learning},
year = {2024},
booktitle = {Proceedings of the 41st International Conference on Machine Learning},
}

@inproceedings{wu2024robust,
  title     = {Robust Heterophilic Graph Learning against Label Noise for Anomaly Detection},
  author    = {Wu, Junhang and Hu, Ruimin and Li, Dengshi and Huang, Zijun and Ren, Lingfei and Zang, Yilong},
  booktitle = {Proceedings of the Thirty-Third International Joint Conference on
               Artificial Intelligence, {IJCAI-24}},
  pages     = {2451--2459},
  year      = {2024},
}

@inproceedings{arpit2017closer,
author = {Arpit, Devansh and Jastrzundefinedbski, Stanis\l{}aw and Ballas, Nicolas and Krueger, David and Bengio, Emmanuel and Kanwal, Maxinder S. and Maharaj, Tegan and Fischer, Asja and Courville, Aaron and Bengio, Yoshua and Lacoste-Julien, Simon},
title = {A closer look at memorization in deep networks},
year = {2017},
booktitle = {Proceedings of the 34th International Conference on Machine Learning - Volume 70},
pages = {233–242},
}

@inproceedings{han2018co,
author = {Han, Bo and Yao, Quanming and Yu, Xingrui and Niu, Gang and Xu, Miao and Hu, Weihua and Tsang, Ivor W. and Sugiyama, Masashi},
title = {Co-teaching: robust training of deep neural networks with extremely noisy labels},
year = {2018},
booktitle = {Proceedings of the 32nd International Conference on Neural Information Processing Systems},
pages = {8536–8546},
}

@inproceedings{qian2023robust,
author = {Qian, Siyi and Ying, Haochao and Hu, Renjun and Zhou, Jingbo and Chen, Jintai and Chen, Danny Z. and Wu, Jian},
title = {Robust Training of Graph Neural Networks via Noise Governance},
year = {2023},
booktitle = {Proceedings of the Sixteenth ACM International Conference on Web Search and Data Mining},
pages = {607–615},
}

@article{wolpert1992stacked,
  title={Stacked generalization},
  author={Wolpert, David H},
  journal={Neural networks},
  volume={5},
  pages={241--259},
  year={1992},
}

@article{akopovadetecting,
  title={Detecting Ethereum Mixers},
  author={Akopova, Amanda},
  year={2024}
}

@inproceedings{perozzi2014deepwalk,
author = {Perozzi, Bryan and Al-Rfou, Rami and Skiena, Steven},
title = {DeepWalk: online learning of social representations},
year = {2014},
booktitle = {Proceedings of the 20th ACM SIGKDD International Conference on Knowledge Discovery and Data Mining},
pages = {701–710},
}

@inproceedings{grover2016node2vec,
author = {Grover, Aditya and Leskovec, Jure},
title = {node2vec: Scalable Feature Learning for Networks},
year = {2016},
booktitle = {Proceedings of the 22nd ACM SIGKDD International Conference on Knowledge Discovery and Data Mining},
pages = {855–864},
}

@article{kipf2016semi,
  title={Semi-supervised classification with graph convolutional networks},
  author={Kipf, Thomas N and Welling, Max},
  journal={arXiv preprint arXiv:1609.02907},
  year={2016}
}

@article{velickovic2017graph,
  title={Graph attention networks},
  author={Velickovic, Petar and Cucurull, Guillem and Casanova, Arantxa and Romero, Adriana and Lio, Pietro and Bengio, Yoshua and others},
  journal={stat},
  volume={1050},
  year={2017}
}

@article{xu2018powerful,
  title={How powerful are graph neural networks?},
  author={Xu, Keyulu and Hu, Weihua and Leskovec, Jure and Jegelka, Stefanie},
  journal={arXiv preprint arXiv:1810.00826},
  year={2018}
}

@inproceedings{yu2019does,
  title={How does disagreement help generalization against label corruption?},
  author={Yu, Xingrui and Han, Bo and Yao, Jiangchao and Niu, Gang and Tsang, Ivor and Sugiyama, Masashi},
  booktitle={Proceedings of the 36th International Conference on Machine Learning},
  pages={7164--7173},
  year={2019},
}

\end{document}